\documentclass[preprint, groupedaddress]{revtex4-1}

\usepackage{hyperref}
\usepackage{epsfig}
\usepackage{graphicx}
\usepackage{subfigure}
\usepackage{latexsym}
\usepackage{color}
\usepackage{fullpage}
\usepackage{dcolumn}
\usepackage{bm}
\usepackage{ulem}
\RequirePackage[cdot,textstyle,amssymb]{SIunits}%
\usepackage{units}
\RequirePackage{xspace}

\newcommand{\Revision}[1]{\textcolor{black}{#1}}

\newcommand{\Mco}{{\ifmmode{M_\mathrm{Co}}\else{$M_\mathrm{Co}$}\fi}}%
\newcommand{\Mpy}{{\ifmmode{M_\mathrm{Py}}\else{$M_\mathrm{Py}$}\fi}}%
\newcommand{\Aco}{{\ifmmode{A_\mathrm{Co}}\else{$A_\mathrm{Co}$}\fi}}%
\newcommand{\Apy}{{\ifmmode{A_\mathrm{Py}}\else{$A_\mathrm{Py}$}\fi}}%
\newcommand{\tco}{{\ifmmode{t_\mathrm{Co}}\else{$t_\mathrm{Co}$}\fi}}%
\newcommand{\tpy}{{\ifmmode{t_\mathrm{Py}}\else{$t_\mathrm{Py}$}\fi}}%
\newcommand{\tcu}{{\ifmmode{t_\mathrm{Cu}}\else{$t_\mathrm{Cu}$}\fi}}%
\newcommand{\tisoe}{{\ifmmode{t_\mathrm{isoE}}\else{$t_\mathrm{isoE}$}\fi}}%

\newcommand{\threedwall}{Landau DW\xspace}%
\newcommand{\threedwalls}{Landau DWs\/\xspace}%

\definecolor{olivier_comment}{rgb}{0.2,0.2,0.8}
\definecolor{olivier_add}{rgb}{0.8,0.2,0.2}
\definecolor{olivier_remove}{rgb}{0.5,0.5,0.5}

\graphicspath{{fig/}}

\begin{document}

\title{Third type of domain wall in soft magnetic nanostrips}

\author{V. D.~Nguyen,$^{1,2}$ O.~Fruchart,$^{1,2}$ S.~Pizzini,$^{1,2}$ J.~Vogel,$^{1,2}$ J.-C.~Toussaint,$^{1,2}$ and N.~Rougemaille$^{1,2,*}$}

\affiliation{$^1$CNRS - Institut N\'{E}EL - F-38042 Grenoble - France \\$^2$Universit\'{e} Grenoble Alpes - Institut N\'{E}EL - F-38042 Grenoble - France \\$^*$nicolas.rougemaille@neel.cnrs.fr}

\date{\today}

\maketitle

\textbf{Magnetic domain walls (DWs) in nanostructures are low-dimensional objects that separate regions with uniform magnetisation. Since they can have different shapes and widths, DWs are an exciting playground for fundamental research, and became in the past years the subject of intense works, mainly focused on controlling, manipulating, and moving their internal magnetic configuration. In nanostrips with in-plane magnetisation, two DWs have been identified: in thin and narrow strips, transverse walls are energetically favored, while in thicker and wider strips vortex walls have lower energy. The associated phase diagram is now well established and often used to predict the low-energy magnetic configuration in a given magnetic nanostructure. However, besides the transverse and vortex walls, we find numerically that another type of wall exists in permalloy nanostrips. This third type of DW is characterised by a three-dimensional, flux closure micromagnetic structure with an unusual length and three internal degrees of freedom. Magnetic imaging on lithographically-patterned permalloy nanostrips confirms these predictions and shows that these DWs can be moved with an external magnetic field of about \unit[1]{\milli\tesla}. An extended phase diagram describing the regions of stability of all known types of DWs in permalloy nanostrips is provided.}

Magnetic domain walls in nanostructures present a range of properties allowing to consider them as objects in their own right. They carry a magnetic charge, have a chirality or a circulation \cite{McMichael1997, Thiaville2012, Chen2013} and can show inertia \cite{Chauleau2010, Thomas2010, Vogel2012}. The domain wall internal structure depends on the thickness, width and material of the constituting magnetic nanostructure \cite{McMichael1997,Nakatani2005, Rougemaille2012}. This internal structure has a strong influence on the DW dynamics, which is of uttermost importance for devices in which the domain wall position is controlled and manipulated by field \Revision{\cite{Ono1999, Thiaville2002, Atkinson2003, Nakatani2003, Beach2005}} or current pulses \cite{Berger1984, Thomas2008, Parkin2008}.

Micromagnetic simulations have proven to be a powerful tool to predict domain wall configurations in confined systems, as well as their dynamic behavior \cite{McMichael1997, Nakatani2005, Thiaville2005}. For flat strips of soft magnetic materials with in-plane magnetisation, like permalloy (Py), simulations showed  the existence of two type of domain walls, the so-called transverse (TW) and vortex (VW) domain walls. A phase diagram was determined, predicting the regions of stability of both walls in a typical thickness range of $\unit[0\mathrm{-}50]{\nano\meter}$ and widths $\unit[100\mathrm{-}600]{\nano\meter}$ \cite{McMichael1997, Nakatani2005}. \Revision{This phase diagram was recently extended to much wider strips and double / triple vortex configurations were found to be low-energy states \cite{Estevez2015}.}

Using micromagnetic simulations and Magnetic Force Microscopy (MFM) measurements, we investigate the magnetic properties of domain walls in the case of Py strips thicker than \unit[50]{\nano\meter}, for which the vortex wall is expected to be energetically favored on the basis of the existing phase diagram \cite{McMichael1997, Nakatani2005, Thiaville2005}. However, above a critical thickness $t_\mathrm{C}\approx\unit[60]{\nano\meter}$, and almost independently of the width of the strip, we find that the vortex wall transforms into a flux closure configuration stretched along the length of the strip, with in-plane magnetisation curling around a Bloch wall of finite length. This prediction is confirmed experimentally by MFM images obtained on a series of 80 nm-thick Py strips of various widths. The observed domain walls show two unusual features: i)~contrary to usual TWs and VWs, they are significantly longer than the width of the strip, ii)~they are characterised by three internal degrees of freedom.

\section*{RESULTS}

\begin{figure*}
\includegraphics[width=\textwidth]{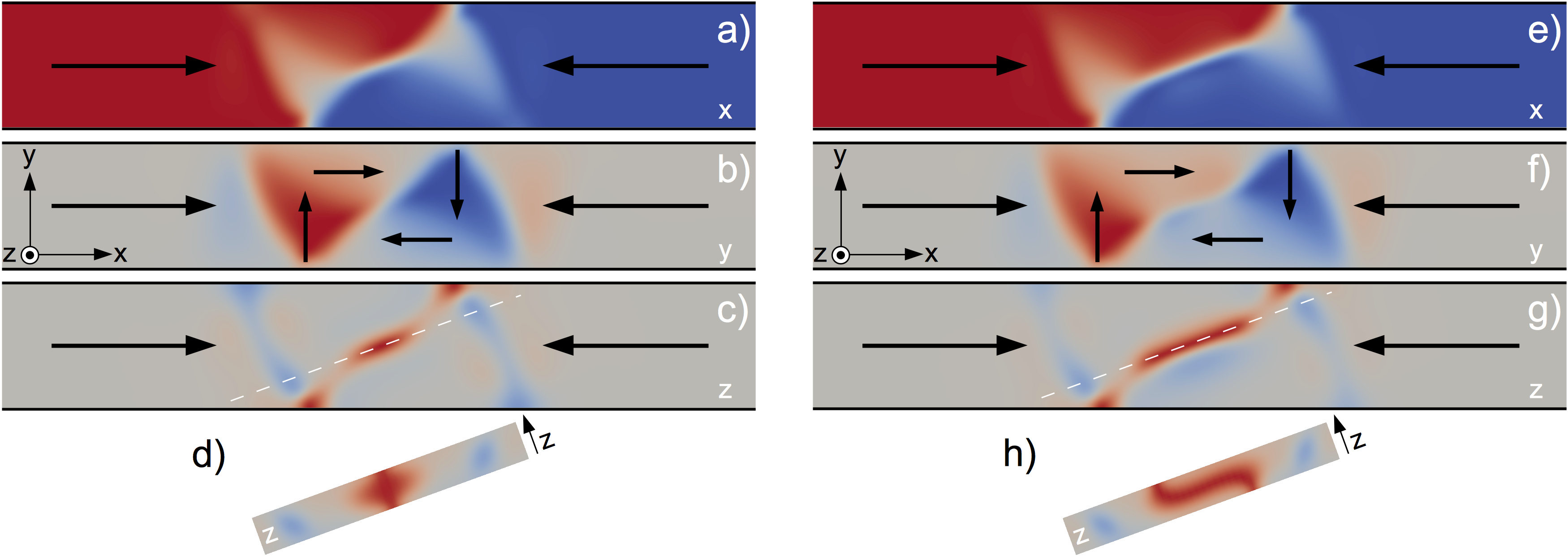}
\caption{\label{fig1} \Revision{(Color online) Micromagnetic configuration of a head-to-head domain wall in a \unit[200]{\nano\meter} wide, \unit[60]{\nano\meter} thick Py strip. (a-d) Vortex domain wall. (e-h) Landau domain wall. In all images, X, Y and Z stand for the length, width and thickness of the strip, respectively. The red / blue color code gives the amplitude of the magnetisation component along X (a,e), Y (b, f) and Z (c, d, g, h). Black arrows represent the local direction of magnetisation. (a-c, e-g) XY cross section views of the domain walls taken at mid-height (z=\unit[30]{\nano\meter}). (d,h) Cross section views of the same domain walls along the dashed white line shown in (c) and (g).}}
\end{figure*}

\Revision{We first consider, numerically, a \unit[200]{\nano\meter} wide, \unit[60]{\nano\meter}-thick Py strip. Two configurations can be obtained, as shown in Fig.~\ref{fig1}: a vortex wall [Figs.~\ref{fig1}(a-d)] and another flux closure magnetic state [Figs.~\ref{fig1}(e-h)]. Contrary to the vortex wall in which the magnetisation distribution can be essentially considered as uniform within the thickness of the strip, the magnetisation distribution in this second DW varies significantly across the thickness. Figs.~\ref{fig1}(d,h)} are cross section views illustrating this distribution: the micromagnetic configuration reveals that the DW is essentially a vortex with a core stretched along the length of the strip, i.e. a flux closure magnetic configuration curling around a Bloch wall instead of a vortex core. The magnetic flux is entering and exiting this Bloch wall at different locations of the domain wall, introducing an asymmetry compared to what is observed in a vortex \Revision{core [see Figs.~\ref{fig1}(d,h)]}. The overall micromagnetic configuration is similar to a rectangular Landau flux-closure pattern \cite{ARR1997, HER1999, Masseboeuf2010}. In the following, we therefore refer to the Landau domain wall when describing this type of DW.

Investigating the influence of the width $w$ of the strip on the critical thickness $t_\mathrm{C}$ at which the \Revision{VW / Landau DW} transition occurs, we find that $t_\mathrm{C}$ is essentially independent of $w$ \Revision{and of the order of $\unit[55]{\nano\meter}$}. This observation, together with the value found for $t_\mathrm{C}$, suggests that this transition may be linked to the N\'{e}el - Bloch wall transition observed in continuous films, which is known to occur at a film thickness of a few tens of nanometers, around \unit[40]{\nano\meter} in Py \cite{Middelhoek1963, Torok1965, Rivkin2009}. Besides, the tilted feature of the Bloch part of the Landau wall is consistent with the known zigzag structure of head-to-head domain walls in thin films with in-plane uniaxial anisotropy \cite{Hubertbook, Hubert1979}. In both systems, the tilt allows spreading the magnetostatic charges over a larger area, and thus decreasing the magnetostatic energy. In strips, the anisotropy is of dipolar origin (shape anisotropy), while it is of magnetocrystalline origin in films.

To check these predictions, we fabricated a series of \unit[40]{\nano\meter} and \unit[80]{\nano\meter}-thick Py strips with different widths, ranging from \unit[200] to \unit[1200]{\nano\meter}. The strips are made with a curved shape to easily set a domain wall in the bent region after the application of an external magnetic field, transverse to the strip \cite{Taniyama1999,Klaui2005}. The resulting magnetic configuration is then imaged after moving the DW in the straight region of the strip. Consistent with micromagnetic simulations, we always observe \threedwalls in the thickest (\unit[80]{\nano\meter}) strips and conventional VWs in the thinnest ones (\unit[40]{\nano\meter}). An example of a MFM image for a \unit[1]{\micro\meter}-wide Py strip is reported in Fig.~\ref{fig2}(a). \Revision{Note, in that image, the black magnetic contrast atop the Bloch wall crossing the entire DW and the unusual length of the overall micromagnetic configuration.}

\section*{DISCUSSION}

\begin{figure*}
\includegraphics[width=\textwidth]{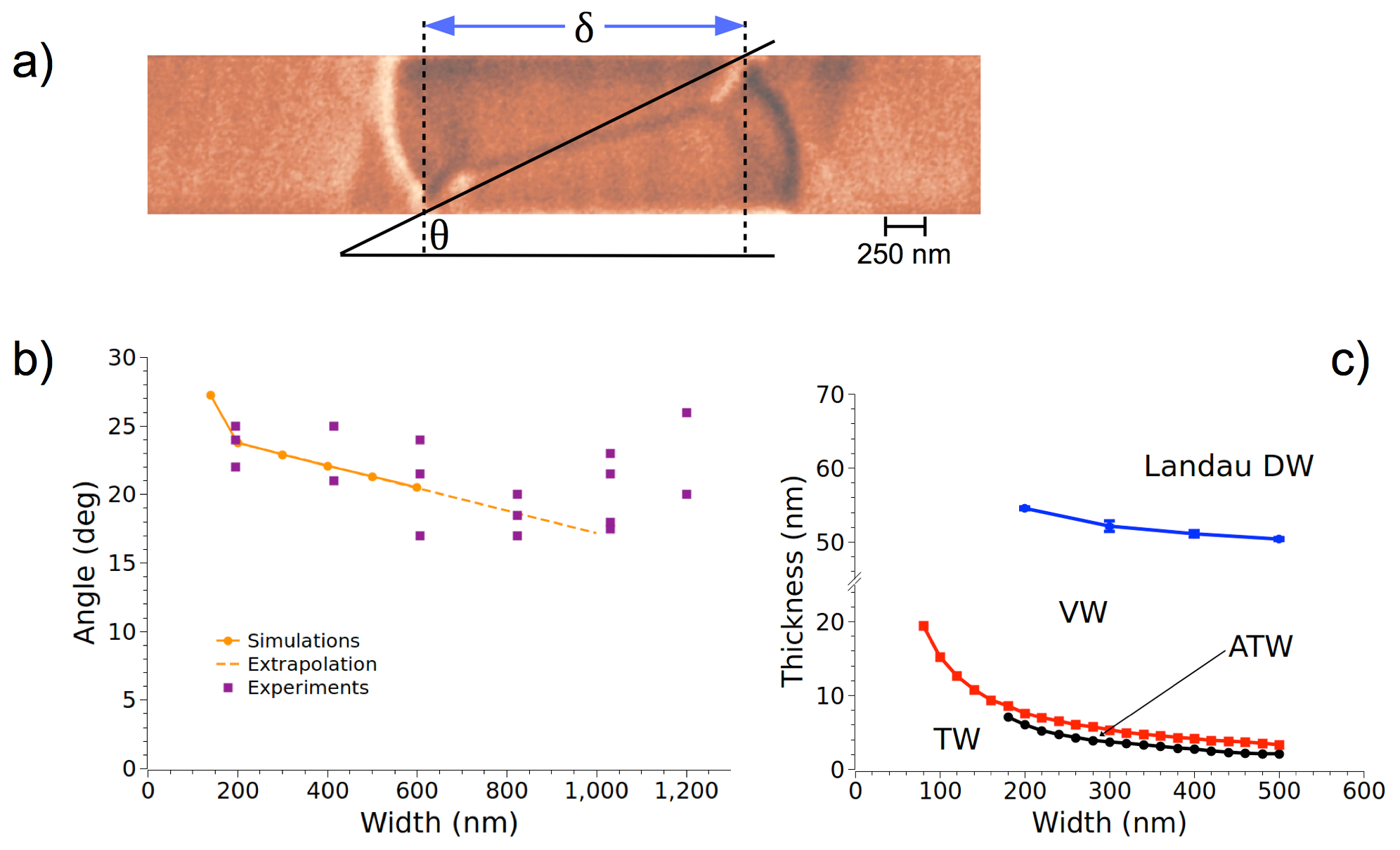}
\caption{\label{fig2} (Color online) (a) MFM image of a head-to-head domain wall in a \unit[1]{\micro\meter} wide, \unit[80]{\nano\meter} thick Py strip. (b) Predicted angle $\theta$  for different width of a \unit[80]{\nano\meter} thick Py strip (yellow dots). The dashed line is an extrapolation of the $\theta$ decrease as the width of the strip becomes larger. The purple squares are experimental data points deduced from the MFM images of \unit[80]{\nano\meter} thick Py strips. (c) Extended phase diagram of Py strips showing the regions where the TW, asymmetric TW (ATW), VW and \threedwall have the lowest energy. The red and black data points are reproduced from Ref.\onlinecite{Nakatani2005}.}
\end{figure*}

When increasing the thickness of the nanostrip, we observe numerically and experimentally an increase of the domain wall width. This result is expected as the amount of volume magnetic charges within the DW increases with the thickness of the strip. Using the MFM images, we measure the angle $\theta$ made by the long axis of the strip and the line passing through the points where the magnetic flux, transverse to the strip, enters \Revision{and exits} the \threedwall[see Fig. 2(a)]. Comparison between the predicted DW angle $\theta$ and the experimental findings, is reported in Fig.~\ref{fig2}(b). \Revision{Although our experimental measurements are scattered, a fairly good quantitative agreement is found between the angle values deduced from the MFM images and the predictions from micromagnetic simulations. The scattering of the experimental data points is likely due to imperfections in the strips, such as edge / interface roughness or pinning sites that could slightly deform the DWs, or can be linked to the way the DWs have been prepared and moved prior to their imaging. However, we want to stress that, above a certain critical thickness (about \unit[50] nm experimentally), we always observe Landau domain walls, independently of the width of the strip (at least in the \unit[200]$\mathrm{-}$\unit[1200] nm range probed in this work). There is no exception and experiments have been reproduced several times on different samples (see Fig.~1 and Fig.~2 in supplementary materials).} \Revision{We also observe} discrepancies for the widest strips (for \unit[1000] 	and \unit[1200] nm-wide strips). We emphasize though that micromagnetic simulations are hardly achievable for such large widths and thicknesses, and understanding if this non-monotonic behavior is an experimental artefact or if it has a physical origin remains an open question. However, micromagnetic simulations performed for Py strips with widths and thicknesses in the range of $\unit[200\mathrm{-}600]{\nano\meter}$ and $\unit[40\mathrm{-}70]{\nano\meter}$, respectively, allow us to provide an extended phase diagram, describing the regions of stability of the TW, VW and \threedwall [Fig.~\ref{fig2}(c)].

\Revision{Interestingly, when the Landau DW is found numerically, it always has a lower energy than the VW. In other words, within the set of parameters we use in the micromagnetic simulations, we never find the Landau DW as a metastable state. For example, a VW is always found if the thickness $t$ of the strip is set to \unit[50]{\nano\meter} or less, even if the initial condition in the micromagnetic simulation is a \threedwall. These results indicate that Landau DWs may not persist as a metastable state and suggest a second-order transition, consistent with the existence of a breaking of symmetry. We emphasize though that characterizing the order of the VW / Landau DW phase transition deserves more investigations, beyond the scope of the present work.}

\begin{figure*}
\includegraphics[width=0.8\textwidth]{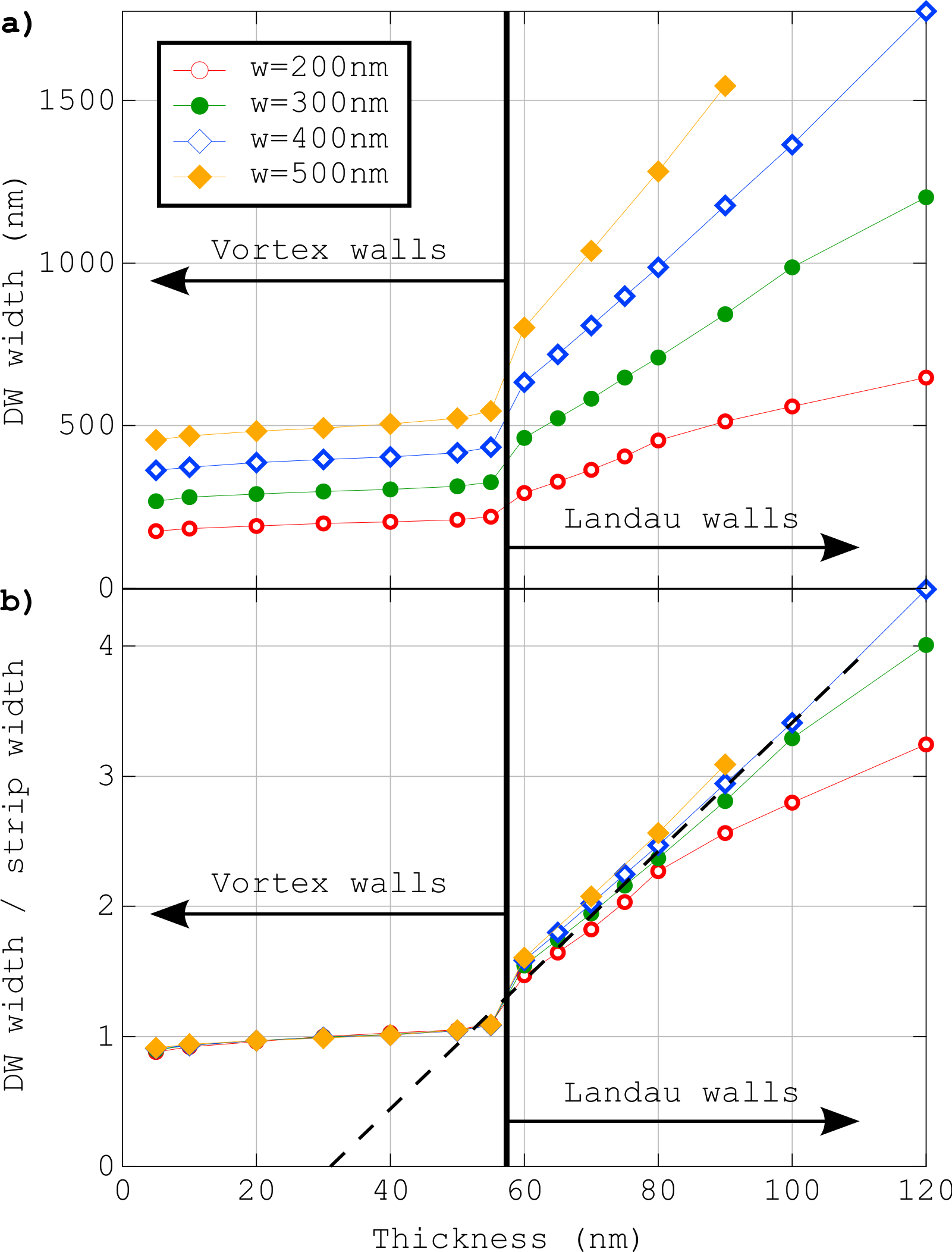}
\caption{\label{fig-scalinglaws} (Color online) Domain wall width $\delta$ (a) and DW width divided by the strip width $\delta/w$ (b) deduced from micromagnetic simulations as a function of the strip thickness~$t$. In (b), the black dotted line is the phenomenological scaling law: $\delta\approx w(t-6\Delta_\mathrm{d})/(4\Delta_\mathrm{d})$.}
\end{figure*}

\begin{figure*}
\includegraphics[width=\textwidth]{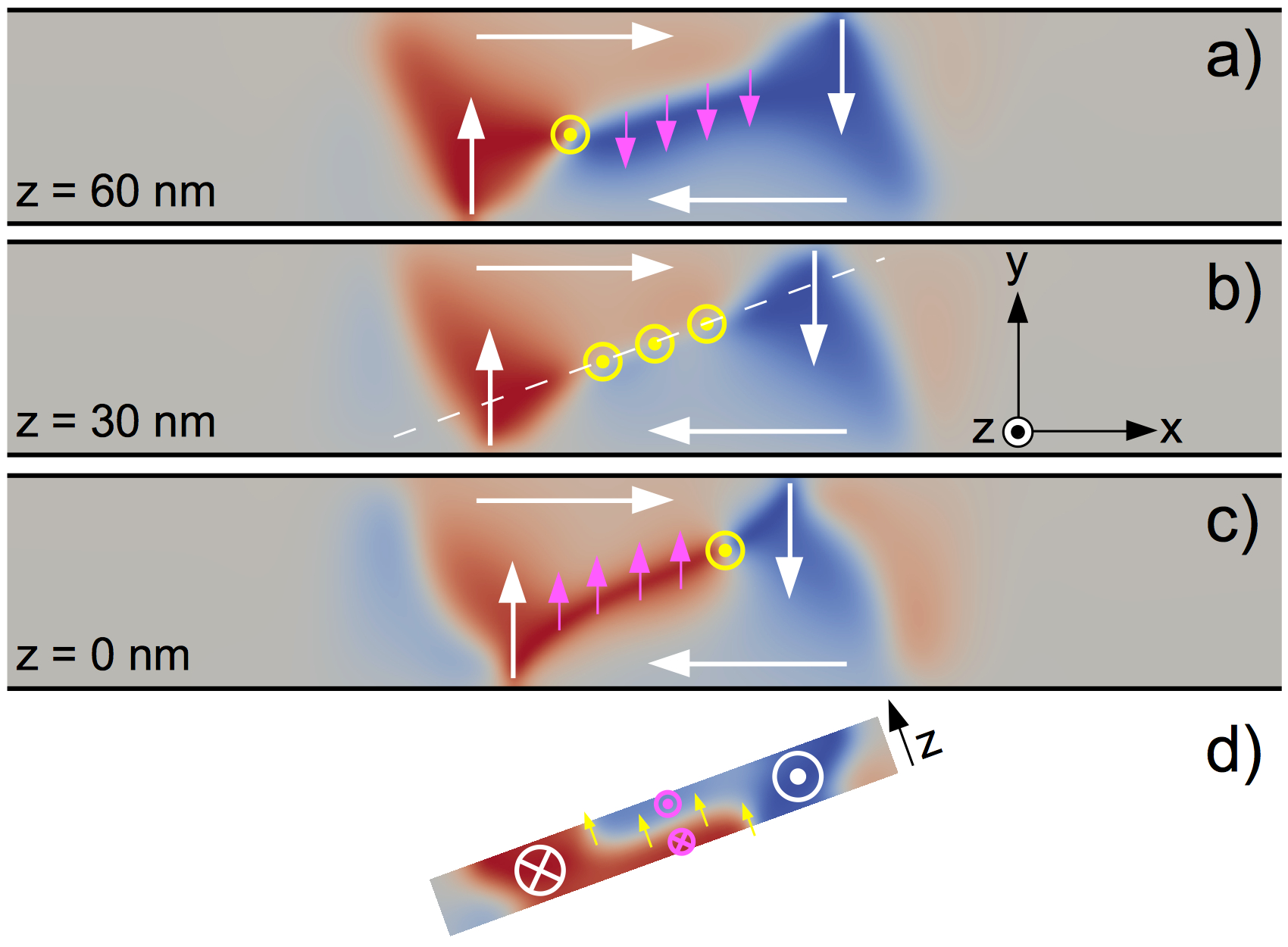}
\caption{\label{fig3} \Revision{(Color online) (a-c) XY cross section views of a \threedwall at different heights [(a) z=\unit[60]{\nano\meter}, (b) z=\unit[30]{\nano\meter}, (c) z=\unit[0]{\nano\meter}] illustrating the 3 magnetic bits: the (clockwise) chirality (white arrows), the polarity of the Bloch wall (yellow arrows) and the direction of the N\'{e}el caps (magenta arrows) for the bottom ($z=\unit[0]{\nano\meter}$) and top ($z=\unit[60]{\nano\meter}$) surfaces. (d) Cross section view along the dashed white line shown in (b). In all images, X, Y and Z stand for the length, width and thickness of the strip, respectively. The red / blue color code gives the amplitude of the magnetisation component along Y.}} 
\end{figure*}

Inspection of the dependence of the DW width~$\delta$ as a function of both the strip width~$w$ and strip thickness $t$ is also useful to compare quantitatively experiments and micromagnetic predictions. In the following, we define the DW width $\delta=w/\tan\theta$  [Fig.~\ref{fig2}(a)]. Figs.~\ref{fig-scalinglaws}(a-b) show $\delta$ and $\delta/w$ as a function of the strip thickness~$t$. The latter clearly reveals two phenomenological scaling laws: $\delta\approx w$ for $t<t_\mathrm{C}$ and $\delta\approx w(t-6\Delta_\mathrm{d})/(4\Delta_\mathrm{d})$ for $t>t_\mathrm{C}$, with $\Delta_\mathrm{d}=\sqrt{2A/\mu_{0}\Mpy^2}$ the dipolar exchange length. The first scaling law simply reflects the double triangular shape of the VW, rather rigidly kept due to the internal $\unit[90]{^\circ}$ walls. The second one may be explained with a crude model accounting for the competition between dipolar and exchange energies, and assuming that the Bloch wall extends along the diagonal of the rectangular shape of the Landau DW. This model predicts $\delta\sim wt$ (see Method section), thus describing accurately the slope of the scaling law deduced from the simulations. The absence in the model of the horizontal shift $6\Delta_\mathrm{d}$ is thought to be linked to the length of the Bloch wall [see Fig.~\ref{fig2}(a)], which is smaller than the full diagonal of the supposed rectangular shape of the Landau pattern. Also, the deviation at large~$t$ for narrow strips results from the ersatz used in the model to estimate the dipolar energy, assuming a thin strip. Qualitatively, the total length of the wall measured experimentally is well reproduced: while transverse and vortex walls have a typical size comparable to the width of the strip \cite{McMichael1997, Nakatani2005}, this \threedwall is significantly larger, \Revision{two - three} times the strip width, consistent with the simulations [Fig.~\ref{fig3}(b)].

\threedwalls have additional interesting properties. Like (asymmetric) transverse and vortex walls, they have two internal degrees of freedom: a chirality~(TW) or circulation~(VW and Landau DW), associated with the clockwise or anticlockwise circulation of the in-plane magnetisation, and a polarity, associated with the out-of-plane direction of the magnetisation within the Bloch wall. But unlike the vortex and transverse walls, the \threedwall has a third internal degree of freedom associated with the in-plane direction of the top and bottom N\'{e}el caps (magenta arrows in Fig.~\ref{fig3}). \Revision{Similar to what is observed in thick, self-assembled Fe micro-dots \cite{Yan2007, Cheynis2009,Fruchart2010}, the two N\'{e}el caps have their magnetisation pointing in opposite directions to minimize the magnetostatic energy of the asymmetric Bloch wall.}

\begin{figure*}
\includegraphics[width=0.5\textwidth]{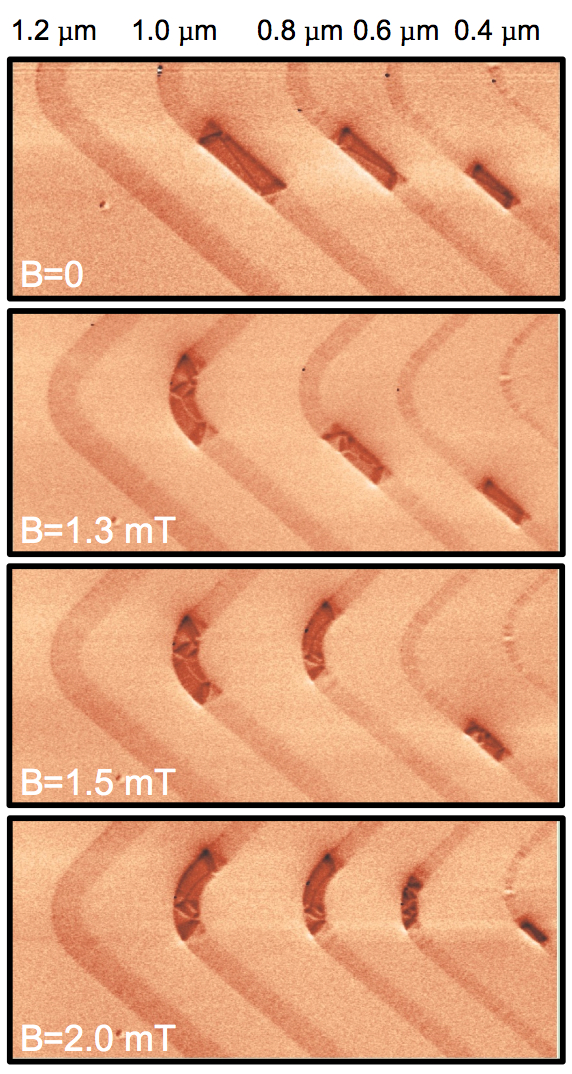}
\caption{\label{fig4} (Color online) MFM images of \threedwalls in \unit[80]{\nano\meter}-thick strips of different widths (indicated in black above the images), before (B=0) and after the application of a $\unit[1\mathrm{-}2]{\milli\tesla}$ magnetic field ($\unit[600]{\milli\second}$ duration) applied along the strip length.}
\end{figure*}

Finally, this entire, three-dimensional micromagnetic configuration can be moved under an applied magnetic field. This is illustrated in Fig.~\ref{fig4}, where a field of 1 mT was applied along the strip length to move the DWs from the straight zone to the bent region of the strip, where they were first nucleated. Similar to what is often observed in thinner Py strips, all \threedwalls could be moved with a relatively small field, 1-2 mT typically. A closer inspection of the MFM images, before and after the application of the external magnetic field, reveals some changes in the internal micromagnetic configurations of the \threedwall. These changes are likely due to the domain wall dynamics during their motion. Understanding the magnetisation dynamics of \threedwalls deserves however more investigations.

\section*{CONCLUSION}

In summary, we have demonstrated experimentally and numerically the existence of a third type of magnetic domain wall in thick Py nanostrips. This wall is characterised by a three-dimensional micromagnetic structure with an in-plane flux closure configuration curling around a Bloch wall of finite length, similar to a Landau flux-closure pattern. The width of this Landau DW is unusually large and can significantly exceed a few microns in Py strips (see Fig.~\ref{fig2} and Fig.~\ref{fig4}). The \threedwall has also interesting properties: besides the chirality of the flux closure configuration and the polarity of the Bloch wall, the wall has well-defined directions for the N\'eel caps. The static and dynamic properties of this wall could open new prospects in nanomagnetism and field- or current-induced domain wall motion.

\section*{METHODS}

Numerically, our approach is based on the finite difference OOMMF code \cite{Oommf}. Material parameters are $\mu_{0}\Mpy=\unit[1.0053]{\tesla}$ for the spontaneous magnetisation, $\Apy=\unit[10]{\pico\joule\per\meter}$ for the exchange stiffness, and zero magnetocrystalline anisotropy. This implies a dipolar exchange length $\Delta_\mathrm{d}=\sqrt{2A/\mu_{0}\Mpy^2}\approx\unit[5]{\nano\meter}$. Although directly applicable to Py, our simulations may be applied to any other magnetic material with no magnetocrystalline anisotropy, scaling lengths with $\Delta_\mathrm{d}$. Cell size in all simulations presented here was set to $\unit[4\times 4\times 5]{\nano\meter\cubed}$. Using a cell size of $\unit[4\times 4\times 2.5]{\nano\meter\cubed}$ yields only negligible differences. Damping is set to~1 to speed up convergence, with no consequence on the result as we are interested in equilibrium states. Magnetic moments at the two extremities of the strips are fixed to avoid non-uniform magnetisation distributions at the edges, and the length of the strips is chosen such that the aspect ratio is at least larger than 10 to limit finite size effects.

Experimentally, the internal micromagnetic configuration of the domain walls gives rise essentially to a monopolar contrast. This is typical for Bloch walls or asymmetric N\'{e}el walls with a significant out-of-plane magnetisation within the core, while N\'{e}el walls would mainly lead to a dipolar contrast. The MFM contrast may be dark or bright in Fig.~\ref{fig2}(a), consistent with the possibility to have either up or down perpendicular component of the magnetisation within the wall [see Fig.~\ref{fig1}(a)].

To determine the width $\delta$ of the Landau domain walls, we use the following model to calculate the dipolar and exchange energies. Dipolar energy of a head-to-head DW (whatever its exact shape) is estimated from a 2D slab carrying the surface magnetic charge $\sigma=2tw\Mpy/(w\delta)=2t\Mpy/\delta$. The resulting dipolar energy integrated in space is $(1/2)\mu_0H_\mathrm{d}^2$, with $H_\mathrm{d}$ of the order of $\sigma/2$ at each surface of the strip and significant only above the wall over a distance $w$. The total dipolar energy is then of the order of $\mathcal{E}_\mathrm{d}\approx(w^2t^2/\delta)\mu_0 \Mpy^2$. The Landau DW being bounded by two segments of wall, nearly perpendicular to the strip whatever the thickness and width (see Fig.~\ref{fig1}), we neglect their contribution in the energy minimization. Most of the remaining energy is then associated with the tilted Bloch wall. The Bloch wall energy per unit length is mostly of exchange origin because of the rather good closure of the flux inside Bloch walls. For the thicknesses considered here, Bloch walls have a close-to-circular cross-section \cite{HUB1969, LAB1969} and thus have an area of the order of $t^2$. This leads to an energy per unit length of $A(\pi/t)^2 t^2=A\pi^2$. Finally, the Bloch wall is assumed to extend over the diagonal of the rectangular shape of the Landau wall, and its energy reads: $\mathcal{E}_\mathrm{W}=A\pi^2\sqrt{\delta^2+w^2}$. Minimization of the total energy versus $\delta$ provides the equilibrium value: $\delta_\mathrm{eq}\sim(\sqrt2/\pi)w t /\Delta_\mathrm{d}$. To obtain this value, we dropped the term related to $w^2$ in the length of the Bloch wall, quickly being negligible above the transition thickness. As often for scaling laws, the numerical factor is not directly applicable, due to the crude approximations made in the model: $\sqrt(2)/\pi\approx1/2$, against $1/4$ in the phenomenological law extracted from the simulations.

\section*{ACKNOWLEDGMENTS}
The authors acknowledge the support of the Nanofab team at the Institut N{\'e}el and also thank S. Le-Denmat for technical help with AFM/MFM measurements.

\section*{Additional information}

Competing financial interests: The authors declare no competing financial interests.

\section*{Author contributions}

V.D.N. and S.P. were in charge of making the samples. V.D.N. and O.F. performed the MFM measurements. J.-C.T. and N.R. made the micromagnetic simulations. V.D.N., O.F. and N.R. analyzed the experimental and numerical data. All the authors have discussed the results. O.F., J.V. and N.R. wrote the manuscript.

\end{document}